\documentclass[10pt,showpacs,amsmath,amssymb,floatfix,superscriptaddress,longbibliography,twocolumn,eqsecnum]{revtex4-2}
\usepackage{amsmath}
\usepackage{physics}
\usepackage{graphicx}
\usepackage[usenames,dvipsnames,svgnames,table]{xcolor}
\usepackage{tcolorbox}
\usepackage[utf8]{inputenc}
\counterwithout{equation}{section}
\counterwithout{figure}{section}
\usepackage[unicode=true,
            pdfusetitle,
            bookmarks=true,
            bookmarksnumbered=false,
            bookmarksopen=false,
            breaklinks=true,
            pdfborder={0 0 0},
            backref=false,
            colorlinks=true,
            hypertexnames=false]{hyperref}
\hypersetup{linkcolor=NavyBlue,urlcolor=NavyBlue,citecolor=NavyBlue}

\begin{document}

\title{Nonlinear dynamics of soliton molecules in a Kerr micro-ring}
\author{Zijian Zhang}
\affiliation{College of Science, Hangzhou Dianzi University, Hangzhou 310018, China}
\author{Chaoying Zhao}
\email{zchy49@163.com}
\affiliation{College of Science, Hangzhou Dianzi University, Hangzhou 310018, China}
\affiliation{State Key Laboratory of Quantum Optics Technologies and Devices, Shanxi University, Taiyuan, 030006,China}
\affiliation{ Zhejiang Key Laboratory of Quantum State Control and Optical Field Manipulation, Hangzhou Dianzi University, Hangzhou, 310018, China}
\date{\today}
\begin{abstract}
The optical Kerr micro-ring provides an ideal platform for the study of dissipative optical solitons. Dissipative solitons are localized waves produced by a precise equilibrium between dispersion and nonlinearity, as well as gain and loss. Dissipative brilliant solitons are vulnerable to external noise, but dissipative dark solitons exhibit greater robustness against noise and losses. This study discusses the division of the input pump light field into transverse electric (TE) and transverse magnetic (TM) modes. $TE$ mode generates bright solitons. $TM$ mode forms dark solitons via cross-phase modulation (XPM), which induces a self-focusing effect. The bright and dark solitons bind into a soliton molecule pair through mutual interactions. We scan the entire detuning interval; within the positive small detuning interval, two modes simultaneously generate bright-dark soliton pair Turing roll states. Within the larger detuning interval, the excited Brillouin scattering noise induces a high suppression ratio (SR) between the dark and bright soliton pumps, which is unfavorable for the formation of the bright-dark soliton pair. The system provides a new multi-soliton manipulation scheme for optical communications. The soliton molecule pairs hold major implications for high-precision measurement and all-optical controlling fields.
\end{abstract}

\maketitle

\section{INTRODUCTION}

Solitons are special solutions of the nonlinear Schrödinger equation (NLSE) \cite{zhu2020modulation}, and have been extensively studied in the contexts of optical communications and ultrafast optics. These unique wave packets maintain a stable shape and constant propagation speed during transmission. When the NLSE includes a negative nonlinear term, it supports spatial bright soliton solutions. Bright solitons typically arise when the input pump field operates in the anomalous group-velocity dispersion (GVD). Conversely, the dynamical behavior of coupled resonators operating in the normal GVD regime has been explored both theoretically and experimentally \cite{sanyal2025nonlinear}, where the input pump field can give rise to dark solitons. The Lugiato-Lefever Equation (LLE) is a specific form of the NLSE, which describes the dynamical behavior of light fields propagating and interacting with micro-resonators. When the pump power exceeds the threshold, the rich nonlinear dynamics based on the Kerr effect, parametric gain, and four-wave mixing process can well explain the generation of the Kerr frequency comb spectrum (KFC). Micro-ring-based KFCs are highly promising as ultra-compact broadband sources for microwave photonic applications. Significant progress has been achieved in KFCs for diverse applications, including photonic radar \cite{trocha2018ultrafast}, dual-comb spectroscopy \cite{pavlov2017soliton}, and low-noise microwave generation \cite{papp2014microresonator}. Modulation instability (MI) plays a critical role in the generation of KFCs by stimulating the growth of sidebands. Quantitative analysis of MI in the anomalous dispersion regime has been theoretically demonstrated in dual-coupled micro-rings \cite{zhu2022frequrncy}. The formation of KFCs in microresonators relies on a delicate balance between dispersion and nonlinear effects. In this process, Kerr nonlinearity induced by cross-phase modulation (XPM) creates phase coupling without energy exchange between the two interacting optical fields.

The analysis of steady-state solutions has demonstrated that KFCs can exhibit various soliton states \cite{zhu2022frequrncy}, including Turing patterns, chaotic states, (breathing) optical solitons, and (breathing) dark pulses. Furthermore, dark-bright soliton pairs have been experimentally demonstrated in both mode-locked lasers and micro-resonators \cite{ning2012bright,shao2015soliton,zhang2022dark}. The aforementioned studies utilized two seed lasers at different wavelengths to jointly pump a microresonator. One laser functions in the bright soliton regime and the other in the dark soliton regime. The XPM-mediated bound states between these solitons allow for the theoretical prediction of bright-dark soliton coexistence across all dispersion regimes—normal \cite{parra2017coexistence}, zero \cite{talla2017existence}, and anomalous \cite{talla2020coexistence} when accounting for third- and fourth-order dispersion effects.

In our paper, an ultrafast laser beam is split into orthogonally polarized $TE$ and $TM$ modes using a  polarization beam splitter (PBS) and subsequently coupled into the optical Kerr micro-ring through an elliptical silicon bus waveguide. In the birefringent microresonator, the $TE$ mode in anomalous dispersion forms bright solitons via self-focusing effects, while the $TM$ mode in the normal dispersion regime generates dark solitons through self-defocusing and XPM effects. Numerical simulations demonstrate that the two solitons can propagate synchronously through group velocity matching. The output spectrum exhibits a symmetric $sech^2$ profile (characteristic of the bright soliton) and a center-concave structure (typical of the dark soliton). This system proposes a novel multi-soliton manipulation scheme, offering potential applications in optical communication and high-precision measurements. Numerical simulations reveal that linked multi-optical fields in the microresonator can elicit substantial nonlinear effects, with XPM acting as the primary coupling mechanism between two orthogonally polarized components. The refractive index alteration generated by coupling displays different values for the two orthogonally polarized components. Moreover, waveguide dispersion is fundamentally influenced by the cross-sectional geometry and dimensions \cite{okawachi2011octave}. A dark-bright soliton pair can be produced using an ultrafast pump light field, a polarizing beam splitter, an elliptical silicon-based bus waveguide, and a birefringent micro-ring resonator.

\section{THEORETICAL MODEL}
Figure 1 illustrates the system setup in which a continuous wave (CW) pump laser operating at $1550nm$ is first divided by a lossless PBS into TE (X-axis/fast axis) and TM (Y-axis/slow axis) polarization modes \cite{zhang2025quad}. The two orthogonally polarized beams are then coupled into a $Si_3N_4$ microresonator using an elliptical silicon-based bus waveguide, where they experience a series of linear and nonlinear processes within the cavity.

\begin{figure}[hp]
    \centering
\includegraphics[width=1\linewidth]{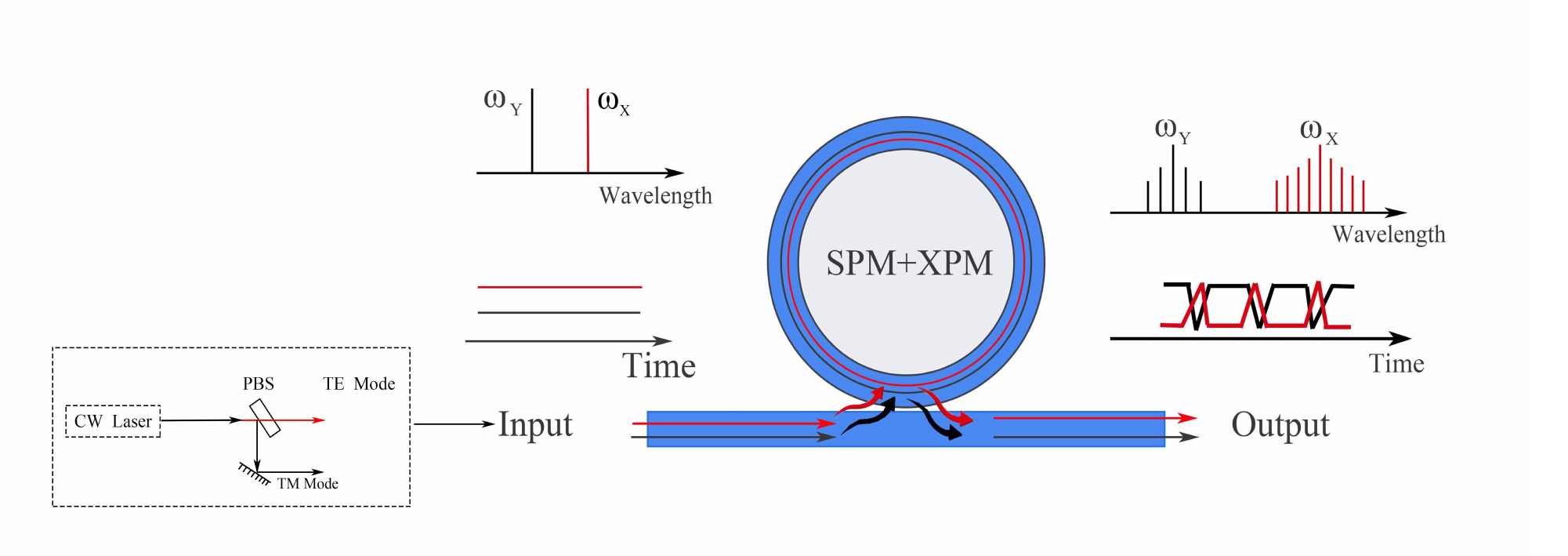}
    \caption{A microresonator houses pairs of dark-bright solitons. The CW with frequencies $\omega_X$ (red) and $\omega_Y$ (black) are injected into the micro-ring, respectively. The self-phase modulation (SPM) and cross-phase modulation (XPM) generate optical frequency combs (OFCs). The output spectra exhibit a symmetric $sech^2$ profile of bright solitons (red) and the central dip characteristic of dark solitons (black).}
\end{figure}

 Here, $\omega_{r_1,r_2}$ represent the microresonator resonant frequencies closest to the pump frequency $\omega_p$. $FSR_{1,2}$ denotes the free spectral range ($FSR$) of the $TE$ and $TM$ modes, respectively. $L$ represents the roundtrip length of the micro-ring. $\delta_{1,2}$ represents the frequency detuning of the $TE$ and $TM$ modes relative to their respective resonance frequencies. These detunings are proportional to the difference between the resonant frequencies and the pump frequency and inversely proportional to the $FSR$. 
\begin{align}
\begin{split}
\delta_{1}=\frac{\omega_{r_1}-\omega_{p}}{FSR_{1} L}
\end{split}
\end{align}
\begin{align}
\begin{split}
\delta_{2}=\frac{\omega_{r_2}-\omega_{p}}{FSR_{2} L}
\end{split}
\end{align}
 
The propagation dynamics of optical pulses in the microresonator are described by coupled LLEs: 
\begin{align}
\begin{split}
\frac{\partial A_{X}}{\partial z}=-\left(\alpha_{1}+i\delta_{1}\right)A_{X}+i\gamma_1\left(\left|A_{X}\right|^{2}+B\left|A_{Y}\right|^{2}\right)A_{X} \\ 
-\beta_{1 X}\frac{\partial A_{X}}{\partial t}-\frac{i}{2}\beta_{2 X} \frac{\partial^{2}A_{X}}{\partial t^{2}}-\frac{1}{6}\beta_{3 X}\frac{\partial^{3}A_{X}}{\partial t^{3}}+A_{Xin}
\end{split}
\end{align}

\begin{align}
\begin{split}
\frac{\partial A_{Y}}{\partial z}=-\left(\alpha_{2}+i\delta_{2}\right)A_{Y}+i\gamma_2\left(\left|A_{Y}\right|^{2}+B\left|A_{X}\right|^{2}\right)A_{Y}\\
-\beta_{1 Y}\frac{\partial A_{Y}}{\partial t}-\frac{i}{2}\beta_{2 Y} \frac{\partial^{2}A_{Y}}{\partial t^{2}}-\frac{1}{6} \beta_{3 Y}\frac{\partial^{3}A_{Y}}{\partial t^{3}}+A_{Yin}
\end{split}
\end{align}

where $z$ is the propagation distance in the cavities. $t$ is the fast time \cite{coen2012modeling}. For $TE$ and $TM$ modes, $\alpha_{1,2}$ represents the intrinsic propagation losses, $\delta_{1,2}$ are the frequency detunings, and $\gamma_{1,2}$ are the nonlinear coefficients. $\beta_{1X, 1Y}$ are the group velocity dispersion (GVD) of the two beams. $\beta_{2X,2Y}$ represents second-order nonlinear dispersion coefficients. $\beta_{3X,3Y}$ represents the third-order nonlinear dispersion coefficients. $A_{Xin}$ and $A_{Yin}$ are the pump optical fields of the two modes after splitting the initial pump light through a beam splitter.

Usually, once a device is fabricated, critical characteristics such as cavity length $L$, intrinsic propagation losses $\alpha_{1,2}$, and frequency detunings $\delta_{1,2}$ become immutable and cannot be adjusted unless assisted by special materials or technologies. For example, integrating PIN diodes into silicon waveguides enables refractive index modification through carrier injection, thereby resolving the issue of disparate dispersions inside a single waveguide. This method solves the problem of different dispersions propagating in a single waveguide. By adjusting the filter parameters, the resonant frequency detuning of the two optical modes in the micro-ring can be independently controlled \cite{miller2015tunable,lu2019deterministic}. Alternatively, tuning the external pump power and the PBS ratio allows for adjustment of the input intensities $A_{Xin}$ and $A_{Yin}$. This power modulation scheme can also be employed to overcome thermal effects in the cavity, thereby stabilizing optical solitons\cite{brasch2016bringing}. Moreover, the fabrication of nanotapers at the waveguide can ensure high coupling efficiency between the waveguide and the input fiber \cite{xu2005micrometre,almeida2003nanotaper}. 

The parameter asymmetry between the $TE$ and $TM$ modes can be expressed by introducing the proportion parameters $\alpha$ and $\sigma$ (see Table 1) and the nonlinear parameter $\Gamma$. 
\begin{equation}
\begin{split} 
 \alpha=\frac{\alpha_1}{\alpha_2},  \gamma_1=\gamma_2=1m^{-1}W^{-1},  \Gamma=\frac{\gamma_1}{\gamma_2}=1, 
 \end{split}
\end{equation}
The loss $\alpha_2$ of the $TM$ mode is used as the normalized reference frame. This normalization scheme is appropriate for analyzing the system with polarization mode competition.  
\begin{table}[h]
\centering
\caption{Normalized parameter definitions}
\label{tab:normalized_parameters}
\begin{tabular}{@{}ll@{}}
\hline
\textbf{Symbol} & \textbf{Original System Expression} \\
\hline
$\zeta$ &  $ z\alpha_2 $ \\
$\tau$ & $ t\sqrt{{2\alpha_2}/{|\beta_{2Y}|}}$ \\
$u$ & $ A_X\sqrt{{\gamma_2}/{\alpha_2}}$  \\
$v$ & $ A_Y\sqrt{{\gamma_2}/{\alpha_2}}$  \\
$\Delta_{1,2}$ & ${\delta_{1,2}}/{\alpha_2}$ \\
$\eta_{1}$ & ${\beta_{2X}}/{|\beta_{2Y}|}$ \\
$\eta_{2}$ & $\mathrm{sgn}(\beta_{2Y})$\\
$\beta_{31}$ & ${\beta_{3X}}/{|\beta_{2Y}|^{3/2}\sqrt{2\alpha_2}}$  \\
$\beta_{32}$ & ${\beta_{3Y}}/{|\beta_{2Y}|^{3/2}\sqrt{2\alpha_2}}$ \\
$A_{Xin,Yin}$  & ${S_{u,v}}$ \\
$\sigma$ & $B\sqrt{\gamma_2/\gamma_1}$ \\
\hline
\end{tabular}
\vspace{2mm}
\end{table}

The group velocity mismatch $d$ between the polarization quantities can be expressed by the following formula
\begin{equation}
\begin{split} 
 d=\frac{\left(\beta_{1X}-\beta_{1Y}\right)\omega_{0}}{2\left|\beta_{2Y}\right|},
 \end{split}
\end{equation}
where $\omega_{0}$ is the optical pulse width. 

Based on Table 1, after normalization, $u$ and $v$ are the steady-state solutions of the two optical pulses. Eq. (3) and Eq. (4) can be rewritten in the normalized form
\begin{equation}
\begin{split}
\frac{\partial u}{\partial\zeta}=-(\alpha+i\Delta _{1})u+i\Gamma(\left|u\right|^{2}+\sigma\left|v\right|^{2})u-\\
d\frac{\partial u}{\partial\tau}-i\eta _{1}\frac{\partial^2 u}{\partial\tau^2}+\beta _{31}\frac{\partial^3 u}{\partial\tau^3}+S_{u},   
\end{split}
\end{equation}
\begin{equation}
\begin{split}
\frac{\partial v}{\partial \zeta}=-(1 +i\Delta _{2})v+i\Gamma (\left|v\right|^{2}+\sigma\left|u\right|^{2})v-\\d\frac{\partial v}{\partial\tau}-i\eta _{2}\frac{\partial^2 v}{\partial\tau^2}+\beta _{32}\frac{\partial^3 v}{\partial\tau ^3}+S_{v}.
\end{split}
\end{equation}
where $\eta_{1,2}$ are the coupling coefficients between the pump light and intra-cavity field. $S_{u,v}$ is the pump amplitude of the $TE$ and $TM$ modes, respectively. This normalization scheme explicitly captures the parameter asymmetry between the two modes via the proportionality parameters $\alpha$ and $\sigma$ and is suitable for analyzing systems with polarization-mode competition. These equations can be numerically integrated along the propagation distance $\zeta$ with the split-step Fourier method \cite{sinkin2003optimization}. When $\alpha\gg 1$, the dominant $TM$ mode corresponds to a low-loss state. When $\sigma>1$, the $TE$ mode exhibits a stronger nonlinearity, making it prone to self-focusing effects. 
 
The MI of the CW steady state plays a crucial role in the formation of frequency combs in the generalized coupled LLE system. In this system, if MI does not occur, the system will quickly settle into the CW steady state. Therefore, we obtain the CW steady state cavity power by setting all the derivatives to zero in Eq. (7) and Eq. (8) and normalizing the nonlinear parameter to $\Gamma=1$:

\begin{equation}
\begin{split}
T=X^{3}-2X^{2}\left(\Delta_{1}-\sigma Y\right) \\
+X\left(\alpha^{2}+\Delta_{1}^{2}-\sigma^{2}Y^{2} +2\sigma Y\Delta_{1} \right),
\end{split}
\end{equation}
\begin{equation}
\begin{split}
R=Y^{3}-2Y^{2}\left(\Delta_{2}-\sigma X\right)+\\ Y\left(1+\Delta_{2}^{2}+\sigma^{2}X^{2} -2\sigma X \Delta_{2}\right).
\end{split}
\end{equation}

where $T=|S_u|^{2}$ and $R=|S_v|^{2}$ represent the pump power of the $TE$ and the $TM$ modes, respectively, and satisfy $T+R=1$. $X=|u|^{2}$ and $Y=|v|^{2}$ denote the intracavity optical powers of the $TE$ and $TM$ modes, respectively. If the cross-phase modulation terms are neglected, Eq. (5) and Eq. (6) reduce to the cubic steady-state equation of a single Kerr cavity \cite{chembo2013spatiotemporal,hansson2013dynamics}. For detunings $\Delta_{1,2}<\sqrt{3}$, the cubic equation is monovaluate, and hence the system is monostable. However, for $\Delta_{1,2}\geq\sqrt{3}$, three solutions exist; the intermediate value is unstable under small perturbations \cite{parra2017coexistence}. Therefore, for the steady-state solutions of a single continuous wave LLE, we typically consider the minimum and maximum values. In our study, the presence of XPM terms containing $\sigma$ leads to more complex solutions. Our primary goal is to generate dark-bright soliton pairs through XPM under varying detuning conditions.

\begin{figure}[hptp]
    \centering   
    \includegraphics[width=1\linewidth]{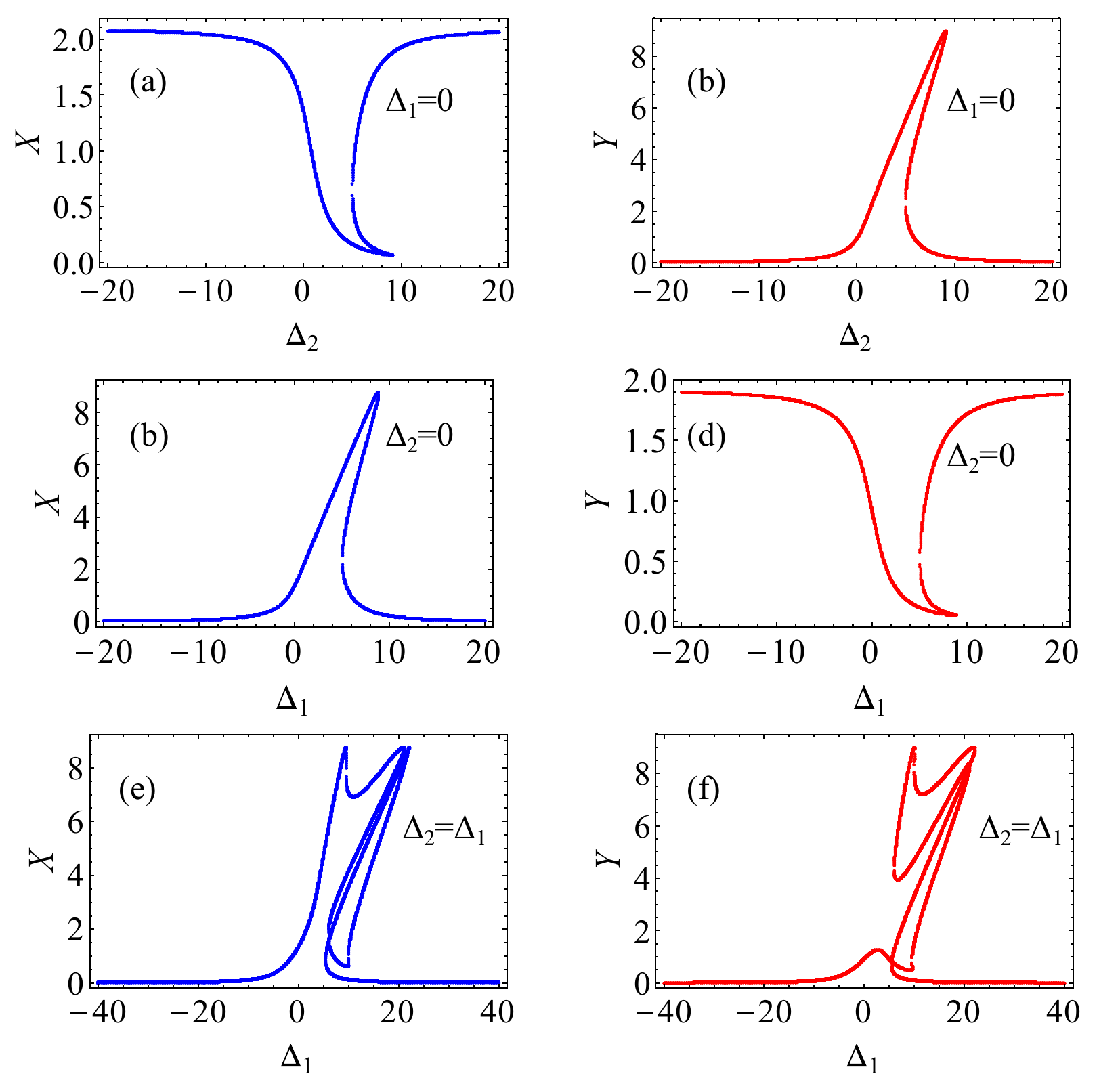}
    \caption{The variation of CW steady-state intra-cavity power with detuning is given by the percentage parameter $\alpha=1.17$ and $\sigma=1.5$, the $TE$ mode initial power $T=12mW$, and the $TM$ mode initial power $R=9mW$. The micro-ring parameters are as follows: radius: $119\mu m$, height: $900nm$, width: $1500nm$, refractive index $n=2$, effective mode area $A_{eff}=0.1\mu m^2$ \cite{bao2019orthogonally}, $FSR=192GHz$. (a) The relationship between $T$ and $\Delta_2$. (b) The relationship between $R$ and $\Delta_2$ with $\Delta_1=0m^{-1}$. (c) The relationship between $T$ and $\Delta_1$. (d) The relationship between $R$ and $\Delta_1$ with $\Delta_2=0m^{-1}$. (e) The relationship between $T$ and $\Delta_1$. (f) The relationship between $R$ and $\Delta_1$ with $\Delta_2=\Delta_1$. The intra-cavity power exhibits a tilted curve due to the Kerr nonlinearity.}
\end{figure}

In Fig. 2(a) and 2(b), we first resolve the detuning of the $TM$ mode at $\Delta_2=0\mathrm{m}^{-1}$. Then, we scan the detuning of the $TE$ mode, $\Delta_1$, from $-20\mathrm{m}^{-1}$ to $20\mathrm{m}^{-1}$, and observe the corresponding steady-state power responses of both modes. As the detuning $\Delta_1$ shifts blue (negative) detuning to the red detuning (positive), the system transitions from low-energy to high-energy states. Notably, when the excitation of $Y$ increases, the pump power $S$ exhibits a pronounced dip, indicating a strong nonlinear XPM effect between the $TE$ and $TM$ modes. Figures. 2(a)-2(d) display four curves that appear as near mirror images of each other, which perfectly illustrate the characteristics of a coupled nonlinear system: each mode competes for resources (power/gain) and modulates each other. In Fig. 2(e) and 2(f), $\Delta_2=\Delta_1$ means the two modes are tuned synchronously; the two detunings are changed simultaneously. 

We further investigate the dynamics by introducing perturbations to the CW steady-state solutions.

\begin{equation}
\begin{split}
u(\zeta,\tau)=u_{0}+\delta u(\zeta,\tau),\\
\delta u(\zeta,\tau)=a_{1} e^{\lambda \zeta+i\Omega\tau}+a_{2}^{*} e^{\lambda^{*} \zeta-i\Omega\tau},\\
v(\zeta,\tau)=v_{0}+\delta v(\zeta,\tau),\\
\delta v(\zeta,\tau)=b_{1} e^{\lambda \zeta+i\Omega\tau}+b_{2}^{*} e^{\lambda^{*}\zeta-i\Omega\tau}.
\end{split}
\end{equation}

Where $a_{1}, a_{2}^{*}, b_{1}, b_{2}^{*}$ are perturbation amplitudes added to two modes, respectively. $\Omega$ denotes spatial mode, $\lambda$ is the wavenumber, and $M$ is a $4\times 4$ matrix.

\begin{equation}
\begin{split}
\mathrm{M}
\left[ \begin{array}{l} 
a_{1} \\ 
a_{2}^{*} \\ 
b_{1} \\ 
b_{2}^{*} 
\end{array} \right] 
= \lambda
\left[ \begin{array}{l} 
a_{1} \\ 
a_{2}^{*} \\ 
b_{1} \\ 
b_{2}^{*} 
\end{array} \right], 
\quad 
\mathrm{M} = 
\left[\begin{array}{llll} 
M_{11} & M_{12} & M_{13} & M_{14} \\ 
M_{21} & M_{22} & M_{23} & M_{24} \\ 
M_{31} & M_{32} & M_{33} & M_{34} \\ 
M_{41} & M_{42} & M_{43} & M_{44} 
\end{array} \right],
\end{split}
\end{equation}

For a single-mode LLE, only the SPM term $M_{12}$ appears. The gain of MI primarily arises from the balance between nonlinear phase modulation and phase mismatch caused by dispersion. For the anomalous dispersion, the dispersion term is given by $D(\Omega)$=$-i\beta_{2}\Omega^2$, which is in phase with the nonlinear term $i|u_0|^2$. This alignment can result in a positive real part of the eigenvalue, thereby enabling MI growth. In contrast, for the normal dispersion, the dispersion term $D(\Omega)$=$-i\beta_{2}\Omega^2$ introduces a negative imaginary part; the same dispersion term introduces a negative imaginary component, which typically suppresses MI and ensures that all eigenvalues have negative real parts. As a result, in single-mode micro-ring resonators, MI occurs only under anomalous dispersion conditions. In the generalized coupled LLEs, each mode is influenced not only by self-phase modulation (SPM, $M_{12}$ and $M_{34}$), but also by cross-phase modulation (XPM, $M_{14}$ and $M_{32}$) from the other modes.

To analyze the MI, we introduce a small perturbation ansatz into the CW steady-state solutions of Eq. (7) and Eq. (8) and linearize the resulting equations. This procedure yields the matrix $\mathcal{M}(\Omega)$, which governs the evolution of the perturbations. The eigenvalue spectrum of $\mathcal{M}(\Omega)$ determines the MI gain as a function of spatial frequency $\Omega$. For each spatial mode $\Omega$, the eigenvalue $\lambda(\Omega)$ is obtained by solving the characteristic equation 
\begin{equation}
\begin{split}
\det(\mathcal{M}(\Omega)-\lambda I)=0.
\end{split}
\end{equation}

In the Fourier domain, we treat $\mathcal{M}(\Omega)$ as an eigenvalue problem. The terms related to the dominant mode can be expressed as:
\begin{equation}
\begin{split}
\mathbf{\mathcal{M}(\Omega)} =
\begin{pmatrix}
\Delta_u(\Omega) & iu_0^2 & i\sigma u_0 v_0^* & i\sigma u_0 v_0 \\
-(u_0^*)^2 & -\Delta_u(-\Omega) & -i\sigma u_0^* v_0^* & -i\sigma u_0^* v_0 \\
i\sigma v_0 u_0^* & i\sigma v_0 u_0 & \Delta_v(\Omega) & iv_0^2 \\
-i\sigma v_0^* u_0^* & -i\sigma v_0^* u_0 & -(v_0^*)^2 & -\Delta_v(-\Omega)
\end{pmatrix},
\end{split}
\end{equation}
where 
\begin{equation}
\begin{split}
\Delta_u(\Omega)=\Delta_1+2|u_0|^2+\sigma|v_0|^2 +D_j(\Omega)+i\alpha,\\
\Delta_v(\Omega)=\Delta_2+2|v_0|^2+\sigma|u_0|^2 +D_j(\Omega)+i\alpha.
\end{split}
\end{equation}
and
\begin{equation}
	D_j(\Omega) = -i\eta_j \Omega^2 + i\beta_{3j} \Omega^3
\end{equation}
is the dispersion term for mode $j=1,2$, accounting for both group velocity dispersion (GVD) and third-order dispersion (TOD). 

The effective nonlinear terms in the pump mode become $i(2|u_0|^2+\sigma|v_0|^2)$, indicating that even if the pump itself lies in the normal dispersion region, the effective nonlinear gain can be significantly enhanced due to the XPM. In the constructed matrix $\mathcal{M}(\Omega)$, although the dispersion term $D_j(\Omega)$ tends to suppress modulation instability, the nonlinear contribution $i(2|u_0|^2 + \sigma |v_0|^2)$ can be sufficiently large to counteract this suppression. As a result, the overall balance may shift toward instability, allowing a positive real part of the eigenvalue to emerge and enabling MI growth.

The MI gain is closely related to the system's CW steady-state solutions, which serve as a precursor to soliton formation. MI tends to occur on the upper branch of the bistable CW solution, while the lower branch remains stable and free of MI, and the middle branch is always linearly unstable. On the upper branch, the system remains linearly stable at small detunings. However, as the detuning increases, MI emerges. For instance, in Fig. 2(a) and (b), if we scan $\Delta_2$ from blue detuning to red detuning, the $TM$ mode power initially resides on the lower branch, crossing the inflection point, and then enters the upper branch. In contrast, the $TE$ mode power begins on the upper branch and then drops to the lower branch. This transition reflects a shift from a uniform CW background to localized solution structures or periodic patterns. Therefore, the steady-state analysis discussed above provides the foundational background for understanding the onset of MI, which originates from the instability of the upper steady-state branch and gives rise to soliton formation.

Figure 3 shows the MI gain calculated on the upper branch. As shown in Fig. 3(a), the $TE$ mode has the abnormal dispersion, while the $TM$ mode has the normal dispersion. It is well known that for a single cavity, MI usually only occurs in the anomalous dispersion region. However, when we transmit simultaneously two orthogonally polarized light beams into a single cavity to generate XPM, MI can occur in the normal dispersion region \cite{zhang2022dark}.

\begin{figure}
    \centering    
    \includegraphics[width=1\linewidth]{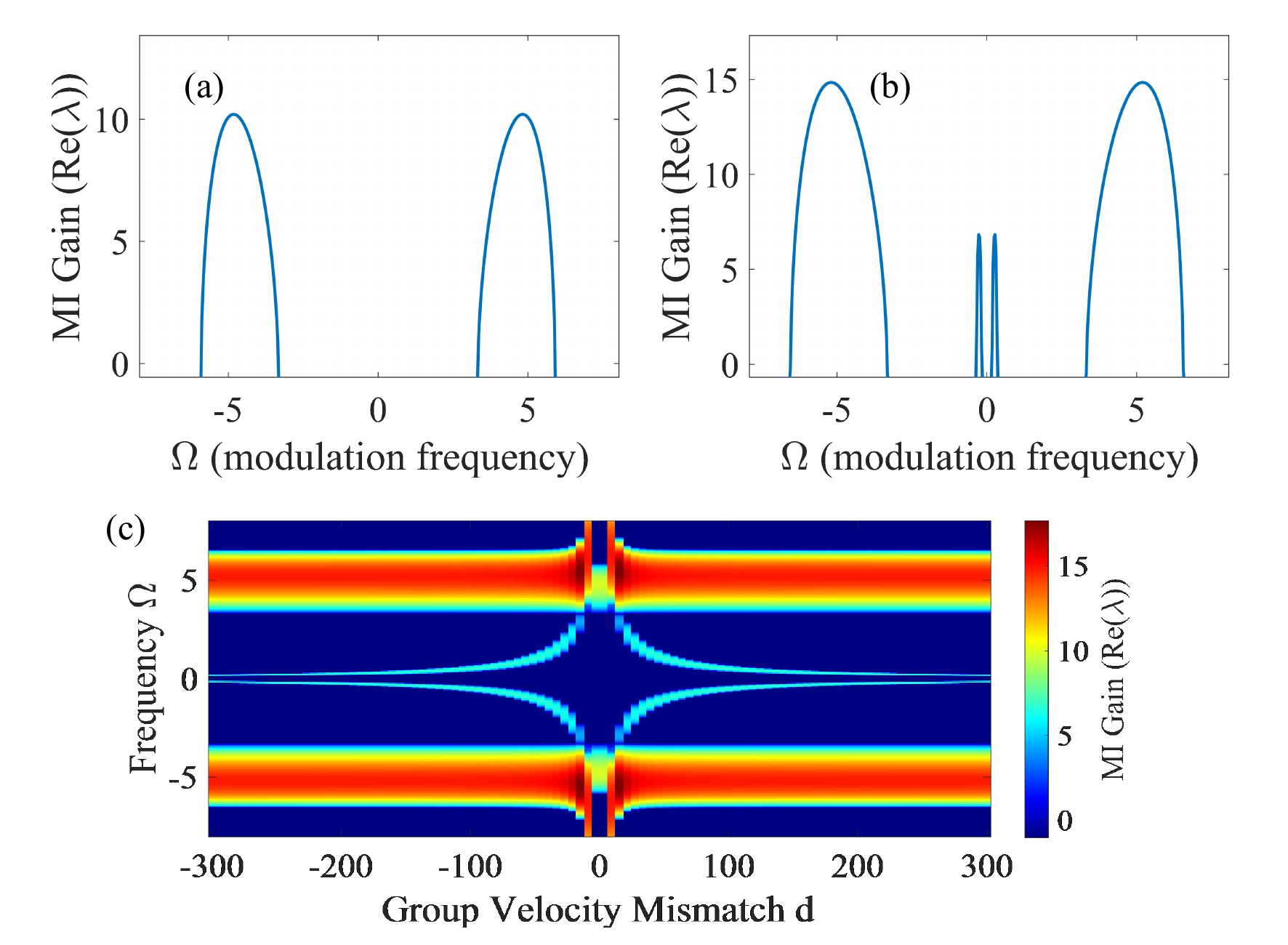}
        \caption{The upper branch of MI gains in the $Si_3N_4$ micro-ring are calculated using the subsequent parameters: $\Delta_1=-2m^{-1}$, $\Delta_2=-12m^{-1}$, $\eta_1=-1m^{-1}$, $\eta_2=1m^{-1}$, and other parameters are consistent with those in Fig. 2. (a) When $d=-0.02fs/m$, the MI gain is computed on the upper branch of the multi-stable state. (b) When $d=-200fs/m$, the increase in $d$ causes the two solitons to evolve independently, resulting in the generation of an extra gain sideband. (c) The relationship graph between d and MI gain spectrum.}
\end{figure}

Our MI analysis indicates that OFCs can be generated when the two modes are in different dispersion regions. However, the MI gain is attributed to the group velocity mismatch $d$ and the XPM. To generate the dark-bright soliton bound state \cite{mao2025temporal}, the $d$ between the two solitons cannot differ too much. If the two orthogonally polarized components propagate with different group velocities, they evolve independently and eventually separate completely in the absence of mutual trapping. Therefore, under otherwise identical parameters, the MI gain spectrum for $d = -200~\mathrm{fs/m}$ shown in Fig. 3(b) exhibits an additional narrow sideband, which arises from XPM. We visualize the influence of $d$ on MI gains in Fig. 3(c). When $d$ is close to zero, the gain spectrum exhibits only two symmetric sidebands, located on either side of the central frequency. However, as $d$ increases, an additional narrow sideband induced by XPM emerges and gradually shifts toward the center frequency. This behavior demonstrates that \(d\) plays a significant role in the MI gain spectrum.

\section{Numerical simulation of dark-bright soliton pairs}

The generalized coupled LLEs indicate that there are paired solitons that maintain their shape invariant through XPM interactions. The dark-bright soliton pairs, which are special solutions of Eq. (7) and Eq. (8). Such solutions should be referred to more accurately as solitary waves \cite{agrawal2000nonlinear}. The group-velocity mismatch represents the most significant hurdle for the existence of XPM-paired solitons. It is possible to realize equal group velocities if the wavelengths of two optical waves are chosen appropriately on opposite sides of the zero-dispersion wavelength such that one wave experiences normal GVD while the other wave lies in the anomalous-GVD region \cite{zhang2022dark}.

Upon achieving the steady state of continuous waves, the $\textit{sech}$ and $\textit{tanh}$ pulses are inserted into the micro-ring as perturbations, leading to the evolution of brilliant and dark solitons. For a dimensionless pulse width of $\omega_0=5fs$, the pulse disturbance produces a brilliant soliton described by $F_{bright}=A_1sech(\omega_0 t)$, with a normalized pulse amplitude of $A_1=0.7fs$. The impulse disturbance produces a dark soliton $F_{dark}=A_2tanh(\omega_0 t)$, where the normalized pulse amplitude is $A_2=0.5fs$ \cite{hu2020dissipative}. The two solitons possess identical breadth and almost equivalent group velocity. Their distinct shapes and amplitudes facilitate coupling via XPM interaction. 

\begin{figure}[h]
    \centering
\includegraphics[width=1\linewidth]{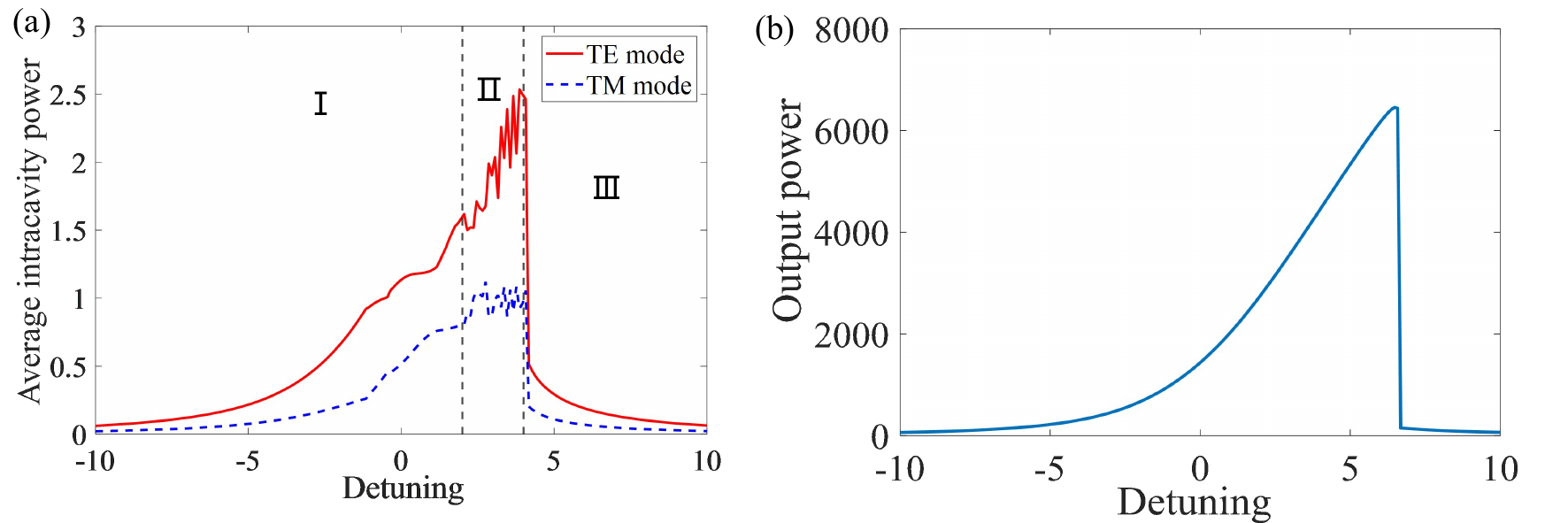}
    \caption{(a) The intra-cavity power variation curves of the $TE$ mode (red solid line) and $TM$ mode (blue dashed line) are shown under different detuning values $\Delta$ after MI. The intensity of the continuous pumping light is $T=6.5mW$, $R=3.5mW$. (b) Plot of detuning versus output power for large detuning.}
    \label{fig:enter-label}
\end{figure}

In Fig. 4(a), the intra-cavity power variation curves of the $TE$ mode and the $TM$ mode under varying detuning $\Delta$ can be divided into three distinct phases. (I): The intracavity power is very low, both the $TE$ and $TM$ modes remain in a CW steady-state. As detuning increases, nonlinear phase accumulation strengthens. (II): The intracavity power reaches a peak. Both modes exhibit step-like transitions and high-frequency oscillations, indicative of a Turing roll regime. With further increase in detuning, chaotic dynamics emerge. In this regime, Turing rolls and frequency combs are generated due to modulation instability. (III): The system leaves the resonance region. The system no longer produces the Turing roll state. The state is dominated by $TE$ mode bright solitons. Due to the strong XPM interaction between the $TE$ mode and the $TM$ mode, the bright solitons absorb the energy from the $TM$ mode, resulting in the suppression of the formation of dark solitons; therefore, the dark-bright soliton pairs cannot be maintained. In Fig. 4(b), there is no oscillatory region, which is characteristic of large-detuning conditions. As the detunings of both modes increase, the system deviates from the resonance, the interaction between the solitons is enhanced, and the driving of the system becomes more and more difficult to energetically sustain the coexistence of multiple solitons, and the system tends toward a single-soliton pair or even a single soliton case.

\begin{figure}[hptp]
    \centering
  \includegraphics[width=1\linewidth]{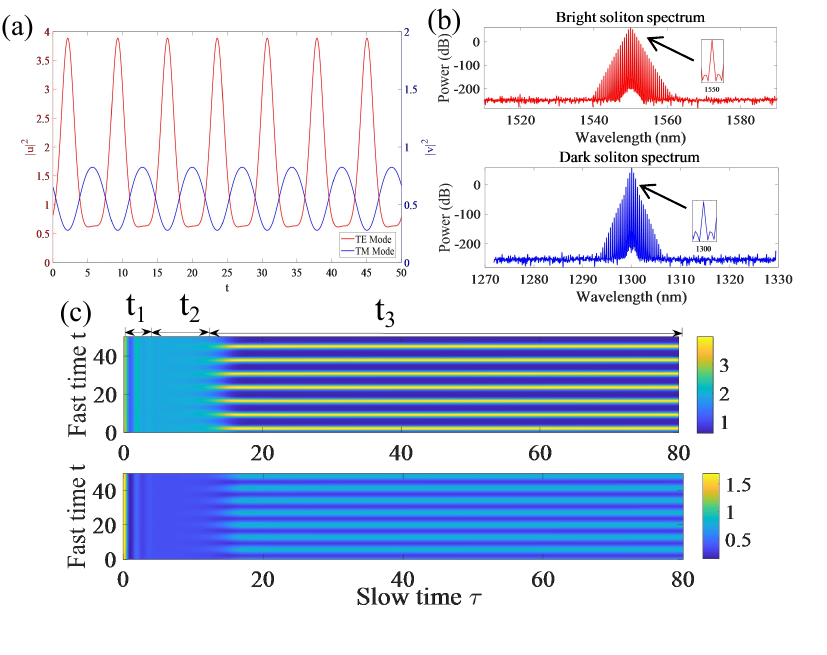}
    \caption{Characteristics of dark-bright soliton pairs in the time domain and frequency domain and evolutionary process with parameters: $\eta_1=-0.5856m^{-1}$, $\eta_2=0.4201m^{-1}$, $\Gamma=1$. Third-order dispersion of the two modes $\beta_{31}=0.2460fs^{3}$, $\beta_{32}=0.2032fs^{3}$. The background field used in Fig. 2 to scan the steady-state solutions corresponding to the detuning $\Delta_1=3m^{-1}$, $\Delta_2=3.5m^{-1}$ and the group velocity mismatch $d=0.017fs/m$. (a) Diagram of Turing roll-state soliton pairs diagram. (b) The Turing comb spectrum with an inset showing the amplified view of the central pump wavelength has a bandwidth of $1nm$. (c) Diagram of the evolutionary spectrum.}
\end{figure}

In Fig. 5, we present characteristics of dark-bright soliton pairs in the time domain, frequency domain and evolution process. Figures. 5(a) and 5(b) depict the intensity distributions of the bright soliton mode $|u|^2$ and the dark soliton mode $|v|^2$ with the fast time $\tau$, respectively. The bright solitons exhibit periodic intensity peaks, while the dark solitons show periodic intensity dips. Moreover, at the peak of each bright pulse, a dark pulse is trapped, forming a localized intensity dip. These results indicate that the coexistence of the two modes in our system can lead to the formation of bound bright–dark soliton pairs. In Fig. 5(b), the bright soliton spectrum is significantly broadened and exhibits relatively high spectral intensity. The inset shows that the bright soliton Turing comb is centered at $1550nm$ and the dark soliton is centered at $1300nm$. In contrast, the dark soliton spectrum also exhibits spectral broadening, but the overall width is narrower and the power is weaker. These spectral characteristics correspond to the sharpness and locality of solitons in the time domain. Figure 5(c) shows the evolution process of solitons over slow time. Unlike dark solitons in normal dispersion micro-ring resonators, the dark soliton here exhibits a smooth waveform without low-intensity oscillations near the pump \cite{yu2022continuum}, its valleys are smooth. The dark–bright soliton pair combines the features of both soliton types, resulting in a unique spectral signature that differs from those of individual solitons.

Subsequently, we investigate a signal dark-bright soliton pair. By adjusting the pump intensity and frequency detuning (primarily the detuning in the multi-stable background field), we find that the pulse energies of the two modes change synchronously throughout the process, reflecting the intrinsic nonlinear dynamical behavior. The gain perturbation simultaneously affects the energies of the two optical pulses, which leads to changes in the interaction potential \cite{zou2025resonantly}. When we driver with different pump intensities, this leads to the formation of supercritical Turing modes, whereupon the Turing roll mode in Fig. 5 disappears. The strong nonlinear coupling between the $TE$ mode and the $TM$ mode results in a stronger energy transfer; the energy of the dark soliton is completely transferred into the bright soliton. Moreover, regardless of the initial perturbation, the final steady-state is always the same. This is similar to the situation in which the initial conditions of the bright soliton led to multi-peak solutions \cite{godey2014stability}. The driven field intensity encompasses not only the pump intensity but also the background field and perturbation intensities.

  \begin{figure}[h]
  \centering
 \includegraphics[width=1\linewidth]{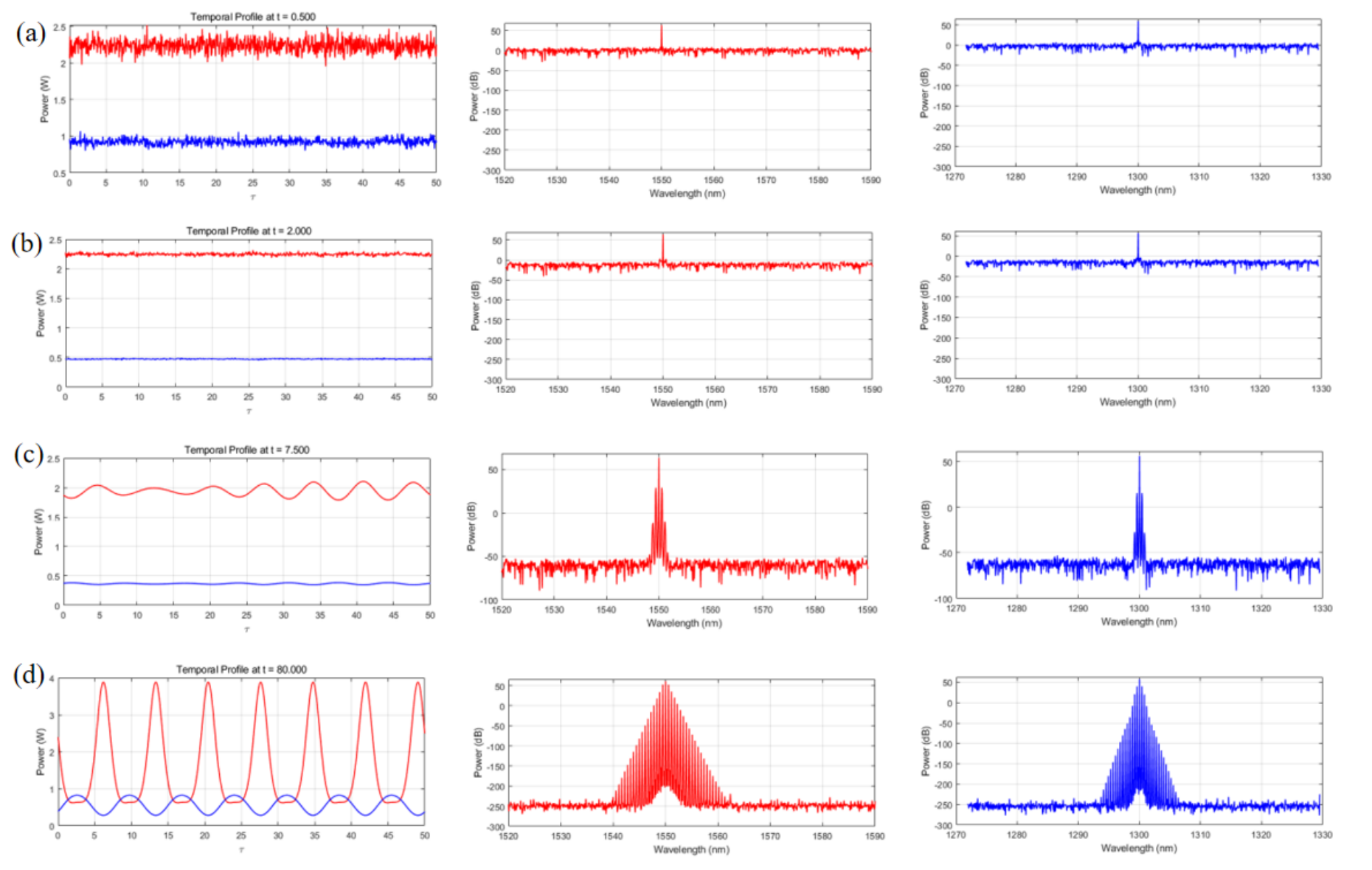}
    \caption{(a), (b), (c), and (d) illustrate the three stages in the evolution diagram of the multiple soliton pairs in Fig. 5. By introducing a perturbed initial state, modulating the unstable process, and ultimately forming the Turing scroll state. Red represents bright solitons, and blue describes dark solitons.}
  \end{figure}

  Figure 6(a) shows the initial light field and spectra following the introduction of the perturbation, where noise predominates, the light in the pumping mode remains unenhanced, and comb spectra are not generated. (b) The system tends to stabilize the uniform background field and has not yet triggered the MI at this time. The spectrum remains focused at the central wavelength, and the bandwidth of the power spectrum is exceedingly narrow. Figure 6(c) corresponds to the beginning of the $t_3$ phase in Fig. 4(c), during which the system starts a structuring process influenced by the MI over time, with significant periodic power fluctuations in the TE mode, which correspond to the spectral bandwidth expansion. The TM mode is influenced by the coupling, resulting in a minor expansion of the spectrum. (d) The system eventually reaches the Turing roll mode \cite{xue2017second}. Multiple equally spaced bright solitons are formed in the TE mode, corresponding to a dense and symmetric OFC structure in the spectrum. The matched arrays of dark solitons are also formed simultaneously in the TM mode, whose spectra have similar periodic broadening characteristics, indicative of the formation of strongly coupled soliton pairs.

\begin{figure}[h]
    \centering
\includegraphics[width=0.8\linewidth]{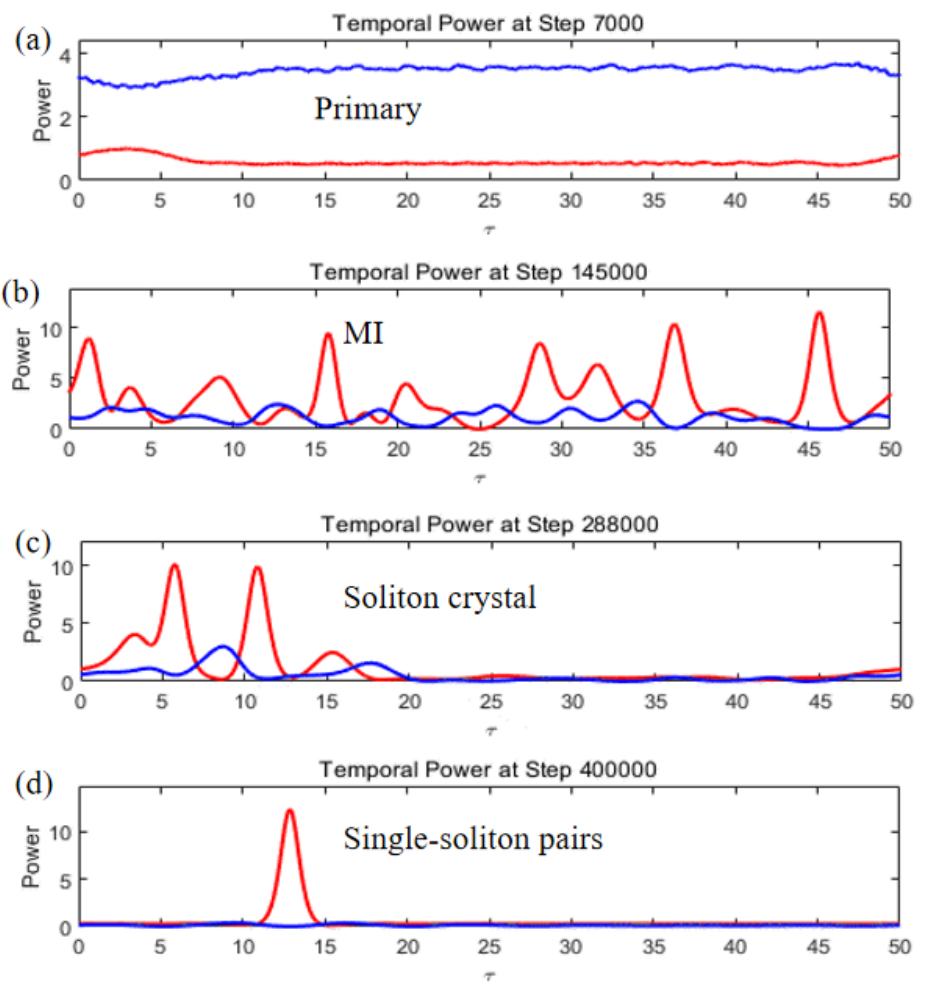}
    \caption{The diagram illustrates the four stages of development for single soliton pairs. Upon modifying the frequency detuning, the perturbed background field (a) shifts from the Turing roll state to the chaotic state. The state of the double soliton pairs remains unchanged. It ultimately stabilizes at pairs of single solitons. (d) Nevertheless, due to the XPM coupling and the SR between the two pumping fields, energy transitions occur from the dark soliton to the bright soliton.}
\end{figure}

In Fig. 7(a), the system enters into the preliminary phase of the multi-soliton lattice. Multiple bright solitons in the $TE$ mode are irregularly arranged, and multiple shallow dark soliton structures appear in the $TM$ mode. The spectrum has not been sufficiently broadened. In Fig. 7(b), the soliton gradually stabilizes, the spacing between the bright solitons in the $TE$ mode tends to be uniform, and the dark solitons in the $TM$ mode become clearer, indicating that the soliton pairs have formed. With the enhancement of nonlinear interactions, the spectrum further broadens. Figure 7(c) shows that the soliton pairs merge or annihilate, resulting in a reduction in quantity, when the system evolves into a lower energy state. The spectrum becomes sparser, the solitons in the $TE$ mode become more concentrated, and the number of dark valleys in the $TM$ mode also decreases simultaneously. In Fig. 7(d), our system finally stabilizes in a single bright-dark soliton pair state. However, due to the strong XPM coupling, the energy of the dark soliton is almost completely transferred into the bright soliton, and to form the stable soliton pairs frequency comb, the spectrum continues to broaden.

It is worth noting that in a stable single dark-bright soliton pair state, the spectra of both the bright and dark modes exhibit cat ear-shape sideband peaks on both sides of the central frequency. This structure stems from the non-flat and asymmetric modulation profile of the soliton pairs in the time domain. Its Fourier transform naturally enhances sidebands. These "cat-ear" peaks not only reflect the localized characteristics and nonlinear coupling of the solitons but also offer a means to adjust the micro-ring frequency comb.
\begin{figure}[h]
    \centering   
    \includegraphics[width=0.8\linewidth]{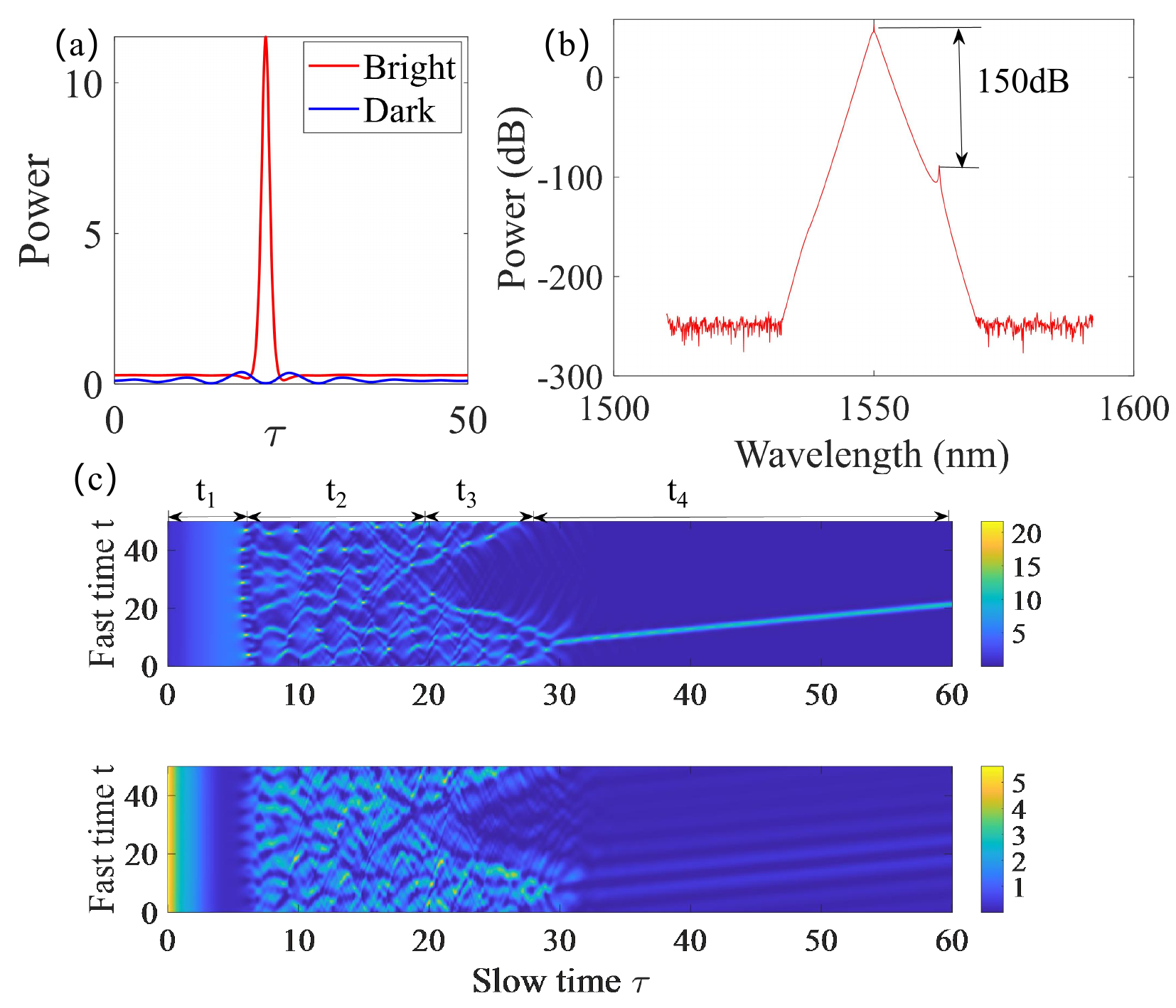}
    \caption{The pump field strengths are $S_u=2.5mW$ and $S_v=1.5mW$, the detunings are $\Delta_2=\Delta_1=5m^{-1}$, and the remaining parameters are identical to those depicted in Fig. 5. Following the occurrence of chaotic states and breather soliton states, the system transitions into a soliton state. However, as a result of the transfer of the energy between bright solitons and dark solitons, the intra-cavity power of the bright soliton mode increases, while the power of the dark soliton mode decreases. (a) Soliton spectrum. (b) The bright soliton spectrum exhibts a 150 dB SR. (c) The entire evolution process of single soliton pairs.}
 \end{figure}
In Fig. 8, it is worth noting that the SR between the dark soliton and the bright soliton pumping reaches an impressive $150dB$, which is due to polarization-dependent suppression \cite{nie2025cross}. The $TM$-polarized pump transfers energy to the $TE$-polarized, resulting in no obvious single dark-bright soliton pairs being observed. The final spectrum diagram is a bright soliton spectrum with a $150 dB$ SR. Fig. 8(b) is the spectrogram corresponding to the Fig. 8(a) soliton spectrum. 

Figure 8(b) reveals the unique dynamics evolution of dark-bright soliton pairs. During the initial phase ($t_1$), the intra-cavity power of the bright soliton gradually increases, while the power of the dark soliton gradually decreases due to resonance offset. In the middle stage ($t_2$), as noise grows, many high-frequency oscillations appear, indicating that our system being in the MI stage. Subsequently, some structures start to merge ($t_3$), which is manifested as the interactions between solitons leading to soliton number reduction and concentration of energy. In the last stage ($t_4$), our system tends to stabilize, eventually forming a single bright soliton. Meanwhile, due to the strong nonlinear coupling, the dark soliton transfers almost all of its energy into the bright soliton. Because of the group velocity mismatch, the soliton trajectories have a certain tilt angle, as shown in Fig. 8(c). The trajectories of the final bright and dark solitons are congruent, with the dark mode trajectory synchronized with the bright mode, verifying the formation of a stable bound state.

\section{CONCLUSION}
In our system, we can obtain the steady-state solutions of continuous waves based on the two normalized coupled LLEs. During this process, due to the XPM coupling, there is a strong modulation between the two polarization states. The solitons of the two polarization states couple to each other to form stable dark-bright soliton pairs. When we study the MI of the continuous wave steady state, we observe that the group velocity mismatch plays an important role. MI also exists in the normal dispersion due to the generation of XPM. Ultimately, we use $sech$ and $tanh$ pulses as the excitation pulses to generate stable soliton pairs. By tuning the pump power and the two detunings in the cavity, it was found that the formation of the soliton pair Turing roll structure was promoted and the Turing comb was generated during the detuning interval of high oscillation. Subsequently, the detuning continues to increase, and there is no longer an oscillatory region throughout the detuning process. The system shifts to a unimodal solution dominated by bright solitons, with the energy-trapping properties of the bright soliton state enhanced, while the dark soliton becomes unstable. The soliton molecular dynamics study mainly focuses on the tuning of the laser parameters and the nonlinear response of the frequency tuning process to explore the potential nature of the soliton intramolecular dynamical system. This extremely sensitive property is anticipated to enable applications in optical communications, optical computing, all-optical control, and related fields.

\begin{acknowledgments}
This work was funded by the State Key Laboratory of Quantum Optics Technologies and Devices, Shanxi University, Shanxi, China (Grants No.KF202503).
\end{acknowledgments}

\bibliography{ref}

\end{document}